# Design and RF Measurements of a 5 GHz 500 kW Window For The ITER LHCD System

J.Hillairet[a], J.Achard[a], Y.S.Bae[b], J.M.Bernard[a], N.Dechambre[c], L.Delpech[a], A.Ekedahl[a], N.Faure[c], M.Goniche[a], J.Kim[b], S.Larroque[a], R.Magne[a], L.Marfisi[a], W.Namkung[e], H.Park[e], S.Park[b], S.Poli[a], K.Vulliez[d],

a. CEA, IRFM, 13108 Saint Paul-lez-Durance, France.
b. National Fusion Research Institute, Daejeon, Korea.
c. PMB/ALCEN, 13790 Peynier, France.
d. CEA, DEN, SDTC, Laboratoire Maestral, Pierrelatte, France.
e. Pohang University of Science and Technology, Pohang, Korea

**Abstract.** CEA/IRFM is conducting R&D efforts in order to validate the critical RF components of the 5 GHz ITER LHCD system, which is expected to transmit 20 MW of RF power to the plasma. Two 5 GHz 500 kW BeO pill-box type window prototypes have been manufactured in 2012 by the PMB Company, in close collaboration with CEA/IRFM. Both windows have been validated at low power, showing good agreement between measured and modeling, with a return loss better than 32 dB and an insertion loss below 0.05 dB. This paper reports on the window RF design and the low power measurements. The high power tests up to 500kW have been carried out in March 2013 in collaboration with NFRI. Results of these tests are also reported.

## 1 - INTRODUCTION

In the current ITER LHCD design, 20 MW Continuous Wave (CW) of Radio-Frequency power at 5 GHz are expected to be generated and transmitted to the plasma. In order to separate the vacuum vessel pressure from the cryostat waveguide pressure, forty eight 5 GHz 500kW CW windows are to be assembled on the waveguides at the equatorial port flange. For nuclear safety reasons, forty eight additional windows could be located in the cryostat section, to separate and monitor the cryostat waveguide pressure from the exterior transmission line pressure. These windows are identified as being one of the main critical components for the ITER LHCD system since first ITER LHCD studies [1] [2] [3] or more recently [4] [5], and clearly require an important R&D effort. In this context and even if the LHCD system is not part of the construction baseline, the CEA/IRFM is conducting a R&D effort in order to validate a design and the performances of these RF windows. In order to begin the assessment of this need, two 5 GHz 500 kW/5 s pill-box type windows prototypes have been manufactured in 2012 by the PMB Company in close collaboration with the CEA/IRFM [6] . The section 2 of this paper reports the RF and mechanical design of a 5 GHz window. Some features of the mechanical design and the experimental RF measurements at low power are reported in section 3. High power results, made in collaboration with NFRI, are detailed in section 4. The development of CW windows is discussed in the conclusion.

## 2 - RF AND MECHANICAL DESIGN

The proposed 5 GHz RF window is based on a pill-box design [2] , i.e. a ceramic brazed in portion of a circular waveguide, connected on either side to a rectangular waveguide section. Typical design rules of thumb of such device are circular section diameter about the same size of the diagonal of the rectangular waveguide (cf. **FIGURE 1**). Without taking into account the ceramic, the circular section length is approximately half a guided wavelength of the circular $TE_{11}$ mode, in order for the device to act as a half-wave transformer. Once optimized, taking into account the ceramic, matching is correct only for a narrow band of frequency and is very sensitive to the device dimensions and the ceramic relative permittivity. The heat losses in the ceramic, which have to be extracted by an active water cooling, depends on the inside electric field topology and of ceramic dielectric loss (loss tangent). Undesirable modes due to parasitic resonances can be excited in the ceramic volume, raising the electric field and such the heat dissipation. This aspect can be decorrelated from the return loss and one can even achieve low return loss but can have high heat losses in the ceramic (>2 kW), as for example in the design proposed in Ref. [4] . So both aspects have to be taken into account during the RF optimization process.

As the main constraints in the design come from the ceramic RF properties knowledge, a sample of BeO ceramic has been ordered from the American Beryllia Company for measuring its permittivity and loss tangent close to 5 GHz. The permittivity and loss tangent values of the sample have been measured at two frequencies around 5 GHz. The design value at 5 GHz has then been taken as the arithmetic mean. Values are reported in Table 1.

**TABLE 1).** RF Characteristics measurements of a BeO sample ordered from American Beryllia.

| Frequency | Relative Permittivity | Loss Tangent |
|---|---|---|
| 4.7 GHz | 6.37 +/- 0.153 | 3.68e-4 +/- 3.73e-5 |
| 5.31 GHz | 6.35 +/- 0.165 | 4.93e-4 +/- 3.91e-5 |
| 5 GHz (arithmetic mean) | 6.36 | 4e-4 |

The final inner dimensions of the 5 GHz window are illustrated in **FIGURE 1**. The circular section is centered on the rectangular one, and dimensions illustrated are thus symmetric with respect to the vertical and the horizontal axis.

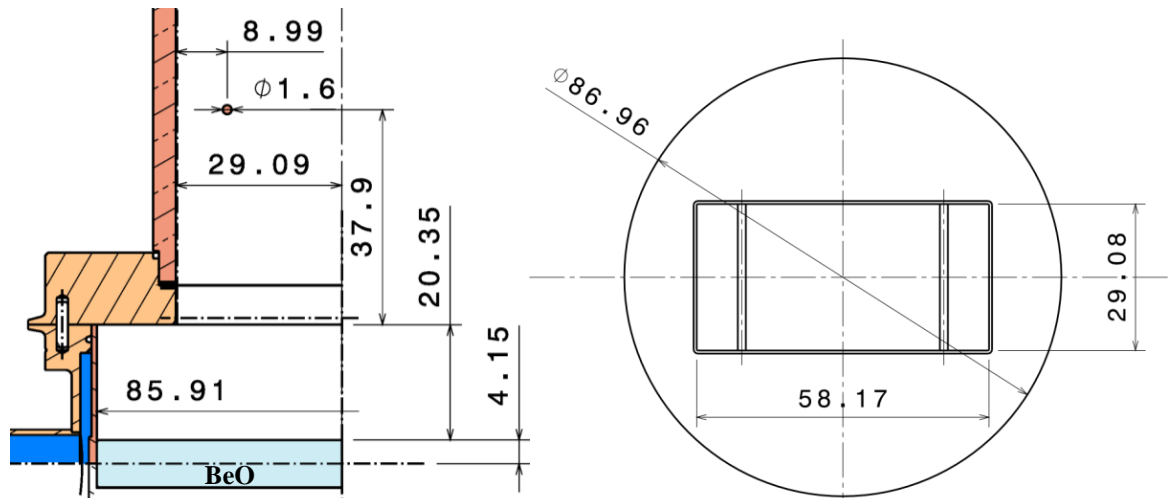

**FIGURE 1).** Final dimensions of the 5 GHz RF window. Left figure: cut drawing (one quarter) with principal dimensions. Right figure: rectangular and circular part dimensions.

### 3 – MECHANICAL DESIGN AND LOW POWER MEASURMENTS

Two RF windows have been manufactured by the PMB Company in collaboration with CEA/IRFM. For ensuring relevant ITER working conditions tests, like high temperature baking up to 240 degC [7] , the window flanges are non-standard and have been designed for both Viton© or Helicoflex© seals. As shown in **FIGURE 2**, a rectangular groove is machined on both sides of the window. Male waveguide adapters are then required to connect the window: these adapters are designed to insure a good electrical contact at the flange by a heel slightly longer than the groove depth. Thus, electrical contact and sealing aspects are mechanically dissociated. In return, additional elements must be connected together, which can lead to additional losses in the transmission line.

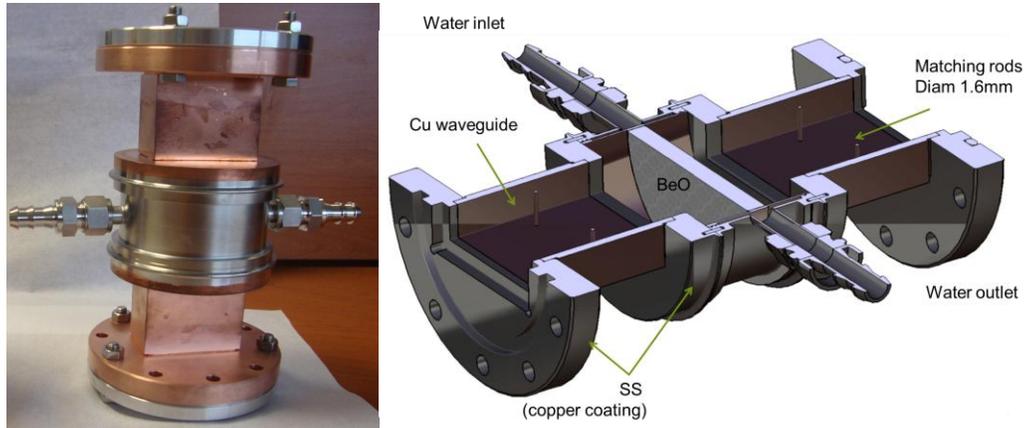

**FIGURE 2).** Left: picture of one window. Right: CAD cut-view of the window.

Low level RF measurements (dBm) are reported in **FIGURE 3**. The return loss ($S_{11}$) at 5 GHz is below -32 dB, within the requested specification and in accordance with the RF modeling. The fact that the return loss of the first window is higher by 3 dB than for the second one is attributed to the mechanical assembly tolerances and the matching rods which were slightly deformed during manufacturing. The insertion losses ($S_{21}$) have been measured between -0.02 dB to -0.04 dB (i.e. 0.46% to 0.92% of input power), both below the requested specification of -0.05 dB at 5 GHz. Precise measurements of the $S_{21}$ are very sensitive to the test bed mechanical assembly, in particular to the quality of the mechanical assembly of all flanges. Thus these measurements reflected not only the window properties but also the part of the assembly for which no calibration was possible, and were then considered conservative.

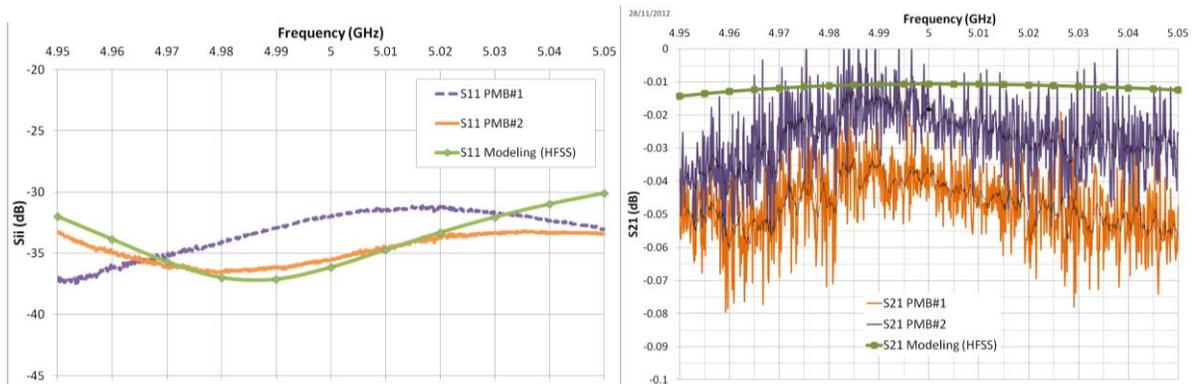

**FIGURE 3).** Return loss (left) and insertion loss (right) for both window. RF modeling prediction in dot-green, from HFSS.

## 4 - HIGH POWER MEASUREMENTS

High power measurements have been performed in collaboration with National Fusion Research Institute in Korea. The setup shown in **Erreur ! Source du renvoi introuvable.** consists in a single window inserted in the transmission line, filled on both side by SF6 gas. The window was cooled with a 8 to 17 L/min water flow at 20-25 degC. Calorimetric measurements of the window water cooling loop were recorded during shots and the temperature of the ceramic was monitored by an infra-red camera. The RF power has been applied during 100 ms pulses starting from 25 kW, and then increased progressively by 25 kW steps up to 250 kW. RF power length has been increased progressively up to 2.5s at 250 kW in more than 30 pulses. The RF power has been increased up to 500 kW by 100 ms pulses, then the pulse length have been increased to few seconds at 500 kW, in more than 50 pulses. A single arc breakdown event has been detected via optical fiber protection system at 250kW, but did not occurred again. The fraction of reflected power measured close to the klystron output ranged from 22.6 dB to 21.6 dB (VSWR 1.16:1 to 1.18:1) for the first window and 21.2 dB to 20.8 dB (VSWR 1.19:1 to 1.2:1) for the second window. An unexpected water temperature increase was measured for both windows ($\Delta T \sim 0.6K$), indicating that between 1.1% to 0.9% percent of the input RF power was lost in the device (ceramic and neighbor waveguides sections), instead of the

0.2% expected from RF modeling. This confirmed the range of values measured at low-power. These over-losses have been found to be mainly located in the ceramic. The infra-red camera measured a central ceramic temperature higher than 250 degC after 3 seconds pulses at 500 kW, far higher that the thermal modeling predictions which was below 100 degC in CW regime. These over-losses are presently explained by a higher dielectric loss than the one measured on a sample. A post-mortem analysis of one of the window ceramic is foreseen in order to confirm this hypothesis. Moreover, the core temperature decrease slope is slower than expected from the thermal modeling, indicating that an unexpected thermal bridge reduce the heat conduction efficiency. Both aspects prevent these two windows prototype to be used in CW regime at full power without exceeding the mechanical stress limits of the ceramic.

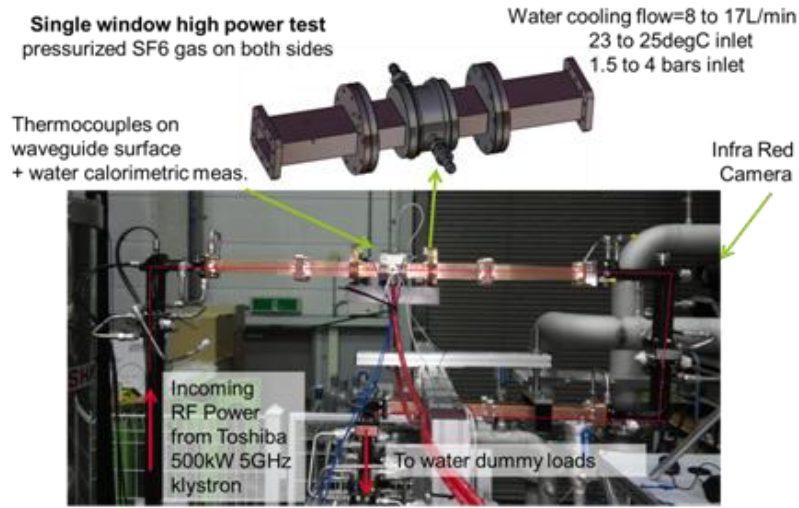

FIGURE 4). High power test setup at 5 GHz/500kW NFRI test bed in Korea.

## 5 – CONCLUSION

Two 500 kW 5 GHz RF windows have been manufactured and tested at low and high power. Low power measurements indicated that both windows were in accordance with the requested specifications, that is, a return loss below 32 dB and an insertion loss below 0.05 dB. However, the high power tests carried out in March 2013 in collaboration with NFRI up to 500 kW / 5 s showed that the losses in the window, mainly located in the ceramic, were five times higher than expected from RF design. A post-mortem analysis of one of the window ceramic is foreseen in order to confirm the hypothesis of a higher dielectric loss than measured on samples before the design of the windows. The mechanical design of future CW prototypes would require a more accurate mechanical assembly (to reduce the assembly tolerances) and of heat removal (prestressing ring placed around the skirt periphery to optimize the radial compressive stress).

## ACKNOWLEDGMENTS

This work, supported by the European Communities under the contract of Association between EURATOM and CEA, was carried out within the framework of the European Fusion Development Agreement. The views and opinions expressed herein do not necessarily reflect those of the European Commission.